# A Well-Behaved Alternative to the Modularity Index

Linton C. Freeman

University of California, Irvine

**1. Introduction.** A major thrust in current network research involves the development of algorithms for uncovering clusters in network data. Network data come in the form of a graph $G = (V, E)$ where $V$ is a set of vertices and $E$ is a set of unordered pairs of vertices, called edges. All these algorithms seek to partition the vertices in $V$ into $m > 1$ subsets within which vertex-vertex connections are dense, but between which such connections are sparse. Social scientists call such subsets *groups*, but to physicists they are *communities* and computer scientists label them *clusters*. But, regardless of the labeling, all these fields are concerned with the same structural form. Here the neutral term *cohesive subsets* will be used to refer to structures of this sort.

None of these cluster-finding algorithms is guaranteed to divide the vertices into subsets that actually have the properties of cohesive subsets. As Newman and Girvan [1] put it:

> Our algorithms always produce *some* division of the network into communities, even in completely random networks that have no meaningful community structure, so it would be useful to have some way of saying how good the structure found is.

An index of the quality of the results of a cohesive subset-finding algorithm must take two properties of the subset into account:

(1) The frequency of *external* edges, those that link pairs of vertices that fall in different subsets.
(2) The frequency of *internal* edges, those that link pairs of vertices both of which fall within a subset.

Such an index should decline in the face of an increasing number of external edges and it should grow in the presence of an increasing number of internal edges. Moreover, it should not be affected by extraneous factors that do not bear directly on these two kinds of edges.

**2. Modularity.** A widely used index of "the quality of a particular division of a network" is Newman and Girvan's *modularity, Q* [1]. $Q$ is focused on the ties that fall within subsets. Let

$k$ = the number of subsets in a graph $G$,

$n$ = the number of edges in $G$,

$n_i$ = the number of edges falling within the $i$th subset,

$e_i = n_i / n,$

$d$ = the sum of the degrees of the vertices in $G$,

$d_i$ = the sum of the degrees of the vertices in the $i$th subset and

$E(d_i) = d_i / d$. Then

$$Q = \sum_{i=1}^{k} (e_i - E(d_i)^2)$$

Here $e_i$ is the fraction of within-subset edges in the network. And $E(d_i)^2$ is the expected fraction of within-subset edges under the assumption that the edges are generated at random, conditional upon maintaining the observed degree distribution of the vertices in $G$.

With respect to external edge frequencies $Q$ performs properly. To see this consider the two-cluster partitioning shown in Figure 1. It is an example of a perfect case of the cohesive subsets we are seeking. No external edges link the two clusters, and all possible internal edges are present within each cluster. For the subsets shown in Figure 1, $Q$ yields a value of 0.50.

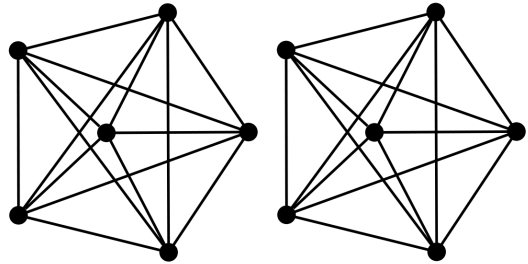

**Figure 1. A Partitioning of a Graph into Two Cohesive Subsets**

Now consider the two cohesive subsets shown in Figure 2. There, a new external edge is displayed—one that crosses between the two clusters. In that case, $Q$ is reduced to 0.468. And that pattern of reduction continues as more external edges are added. Figure 3 shows the decline in values of $Q$ as additional cross-cutting external edges are introduced.

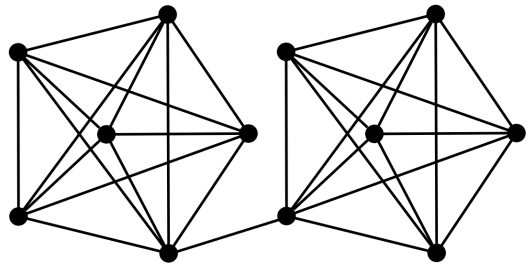

**Figure 2. The Partitioning Shown in Figure 1 Plus an External Edge**

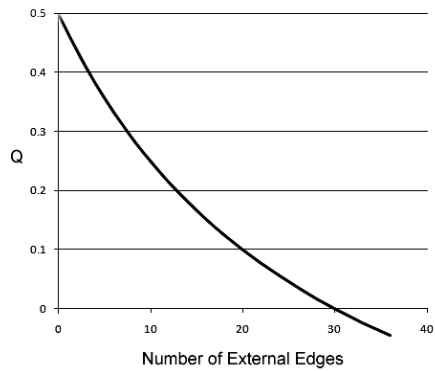

**Figure 3. Declining Values of $Q$ as External Edges Are Added**

*Q*, however, does not perform as well with respect to several other structural properties [2, 3]. Consider first, internal edges. In the perfect cohesive subsets shown in Figure 1, *Q* = 0.50. In that partitioning all the possible internal edges in each cluster are present. Now suppose we remove one edge from each of the clusters as shown in Figure 4. Ideally, we should expect the value of *Q* to decline. However, the value of *Q* for the partitioning shown in Figure 4 is still 0.50. Removing those edges did not decrease its value. Moreover, if we continue to remove edges, systematically, one from each cluster, *Q* continues to produce a value of 0.50 even when there remains only one edge on each side of the split. Although cohesive subset-finding seeks partitions where there are many internal edges, *Q* turns out to be indifferent to the presence or absence of internal edges.

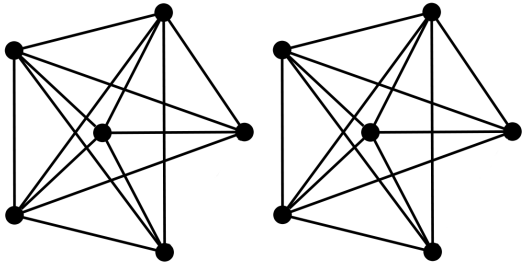

**Figure 4. A Partition into Two Clusters in which an Edge is Missing from Each Cluster**

In addition, the values produced by *Q* are also confounded by the effects of two extraneous factors, subset numbers and edge inequalities. The number of cohesive subsets that are being evaluated constrains the range of values that *Q* can produce. Where *m* is the number of cohesive subsets presented to *Q*, the upper limit that *Q* can reach is (*m* - 1)/*m*. That is why the perfect partitioning displayed in Figure 1 only produces a value of *Q* = 0.50. In contrast, if instead of the two cohesive subsets displayed in that image, there were 100 identical cohesive subsets, then *Q* would take a value of 0.99. In both cases, however, the partitioning is equally flawless.

The impact of inequalities in the numbers of edges contained in the subsets is equally confounding. The value produced by *Q* is reduced to the degree that the clusters with which it is confronted contain differing numbers of edges. Figure 1 and Figure 5, for example, both display perfect partitionings. Both involve graphs that display two cohesive subsets that contain all internal edges and no external edges. But while the data of Figure 1 produce *Q* = 0.50, for the data of Figure 5, *Q* is reduced to 0.14.

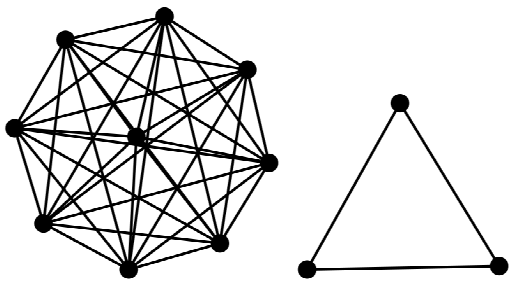

**Figure 5. Two Cohesive Subsets with Unequal Numbers of Edges**

Overall, then, as a tool for evaluating the success of a partitioning, *Q* seems to have some serious limitations. When it is faced with a perfect partitioning, like the ones displayed in Figures 1 and 5, *Q* may still vary between almost 0 and almost 1. It will approach 0 as the differences in the numbers of ties in different clusters become large. It will approach 1 as the number of clusters grows.

**3. Borgatti's $\eta$.** An alternative to $Q$ is presented by an index developed by Borgatti and included in the program Ucinet [4]. $\eta$ is based on the Pearson product-moment correlation. Pearson's correlation is calculated between the observed pattern of edges linking vertices and a perfect partitioning. Let $x_{jk} = 1$ if vertex $j$ and vertex $k$ fall in the same cohesive subset. And let $y_{jk} = 1$ if vertex $j$ and vertex $k$ are adjacent in the data matrix. Then $\eta$ is simply the correlation between the entries in the corresponding cells in the $x$ and the $y$ matrices. $\eta$, then, varies between 1.0 and -1.0. It takes a value of 1.0 whenever all possible internal edges and no external edges are present—whenever it is faced with a perfect partitioning. And it takes a value of -1.0 when no internal edges are present and all possible external edges are present. Thus, $\eta$ measures the degree to which the observed partitioning approaches the ideal.

As was the case with $Q$, $\eta$ declines as external edges are present in $G$. Adding 1 external edge to the graph in Figure 1 yields an $\eta$ of 0.97. Figure 6 illustrates the continuing decline of $\eta$ as edges added to that graph.

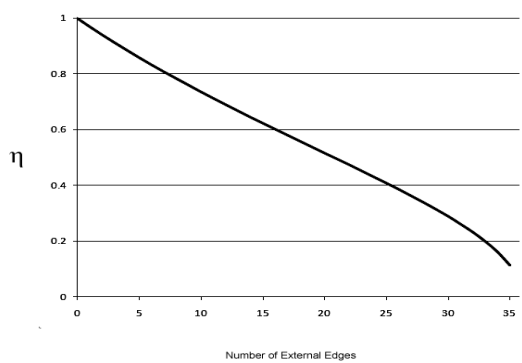

**Figure 6. Declining Values of $\eta$ as External Edges Are Added**

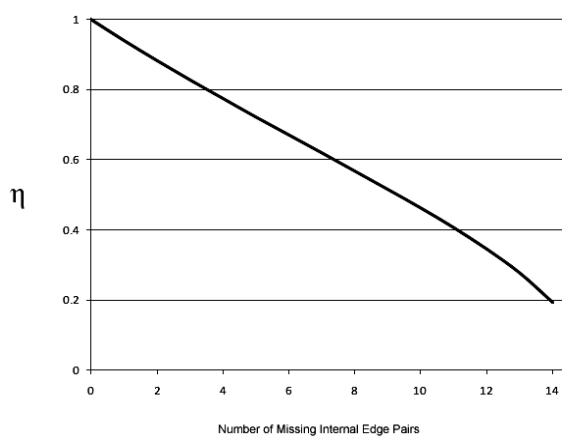

**Figure 7. Declining Values of $\eta$ as Internal Edges Are Removed**

When it comes to missing internal edges $\eta$ fares better than $Q$. While $Q$ was unresponsive to missing internal edges, $\eta$ declines as internal edges are removed. The graph shown in Figure 4, for example, yields a $\eta$ value of 0.97. And as successive pairs of edges are removed, $\eta$ continues to decline. That pattern is shown in

Figure 7.

Unlike *Q*, *η* is not responsive to differences in the number of subsets being evaluated. The perfect partition of Figure 1 yields a *η* of 1.0 and a graph containing 100 cohesive subsets like those in Figure 1 also yields the same value of 1.0. Moreover, *η* does not respond to inequalities in the numbers of edges in different subsets. The perfect partition shown in Figure 5, for example, yields a *η* of 1.0.

Overall, then, *η* is straightforward; it indexes the success of a partitioning by comparing the observed pattern of connections with an ideal pattern. In so doing, it avoids producing the kinds of anomalous results that are produced by *Q*. Unlike *Q*, Borgatti's *η* is not normalized in terms of a set of expectations based upon the degree distribution of *G*. But that seems a small cost to eliminate all the ambiguity produced by *Q*.

## References

[1] M. E. J. Newman and M. Girvan, "Finding and evaluating community structure in networks," *Physical Review E*, vol. 69, no. 026113, 2004.

[2] U. Brandes , D. Delling , M. Gaertler , R. Görke, M. Hoefer , Z. Nikoloski and D. Wagner, "On finding graph clusterings with maximum modularity," *Proceedings of the 33rd International Workshop on Graph-Theoretic Concepts in Computer Science. Lecture Notes in Computer Science*, 2007.

[3] H. Du1, D. R. White, Y. Ren and S. Li , "A normalized and a hybrid modularity ," Draft paper, *eclectic.ss.uci.edu/~drwhite/links2pdf.htm*

[4] S. P. Borgatti, M. G. Everett and L. C. Freeman, *Ucinet for Windows: Software for Social Network Analysis*. Harvard, MA: Analytic Technologies, 2002.